\input amstex
\documentstyle{amsppt}

\hsize=4.75in
\vsize=8in


\def\Phix{{\phi (x)}}
\def\phix{{\varphi (x)}}
 
\def\dz{{{\partial}_{0}}}
\def\dmu{{{\partial}_{\mu}}}  
\def\dnu{{{\partial}_{\nu}}}
\def\dno{{{\partial}^{\nu}}}
\def\dla{{{\partial}_{\lambda}}}
\def\Fmnl{{F_{\mu \nu \, ( \lambda )} (x)}}
\def\jolx{{j_{0 \, ( \lambda )} (x) }}
\def\ga{{\gamma_a}}
\def\gb{{\gamma_b}}
\def\Wa{{W_a}}
\def\Wai{{W_a^{-1}}}
\def\Wb{{W_b}}
\def\Wbi{{W_b^{-1}}}
\def\eab{{\varepsilon_{a b}}}
\def\Uab{{U_{a b}}}
\def\Uabi{{U_{a b}^{-1}}}
\def\Ucd{{U_{c d}}}

\rightheadtext {A Model for Charges of Electromagnetic Type}
\leftheadtext {D. Buchholz, S. Doplicher, G. Morchio, J.E. Roberts, F.
Strocchi}
\topmatter
\title
A Model for Charges of Electromagnetic Type
\endtitle
\author
Detlev Buchholz, Sergio Doplicher, Gianni Morchio,\\ John E. Roberts,
Franco Strocchi
\endauthor
\affil
Institut f\"ur Theoretische Physik der Universit\"at G\"ottingen,\\
D-37073 G\"ottingen, Germany\\
Dipartimento di Matematica, Universit\`a di Roma ``La Sapienza'',\\
I-00185 Roma, Italy \\
Dipartimento di Fisica dell'Universit\`a, I-56126 Pisa, Italy,\\ 
Dipartimento di Matematica, Universit\`a di Roma ``Tor Vergata'',\\
I-00133 Roma, Italy\\ 
Scuola Normale Superiore, I-56126 Pisa, Italy
\endaffil
\endtopmatter
\document
\heading
1. Introduction
\endheading

Despite its successes, the theory of superselection sectors still
needs extending if it is to cover all cases of physical interest. Even
if we restrict our attention to four spacetime dimensions, the two
cases that have been treated successfully, localized charges
[1] and cone--localized charges [2], do not include
the physically relevant class of theories with a local Abelian gauge
group, notably quantum electrodynamics.

The insights into the superselection
structure of the latter theories gained so far, basically have the
character of 
no--go--theorems. We recall in this context that, as a consequence of
Gauss'
law, electrically (or magnetically) charged superselection
sectors\smallskip 
\item{(a)} are not invariant under Lorentz transformations
[3], [4], \item{(b)} cannot be generated from the vacuum by field
operators localized in bounded spacetime regions [1], [5]
or arbitrary spacelike cones [6], and \item{(c)} do not admit a
conventional particle interpretation (infraparticle problem)
[3], [4]. \smallskip
Consequently, it does not seem possible to characterize these
super\-se\-lection sectors by the spectrum (dual) of some group of
internal symmetries nor to assign to each sector some definite
statistics. Whereas in massive theories, this is always possible
[7].

  The aim of the present article is to discuss a simple
but instructive model showing that the situation is actually not as
hopeless
as it may seem. Even though Gauss' law
holds for the charged states in this model with all its inevitable
consequences, the corresponding sectors can be labelled by the
spectrum of some internal symmetry group and have well defined
statistics. More interestingly, the properties of these charged states
seem to point to a general argument allowing one to
establish these features for all charges of electric and magnetic type. 
Our present findings thus set the stage for a more
general analysis of this issue to be carried out elsewhere.

   Prior to discussing our model, let us explain in
heuristic terms why these rather unexpected results are in harmony
with the above no--go--theorems. First, the Coleman--Mandula Theorem 
apparently says that the charged superselection sectors cannot be 
characterized by an internal symmetry group for it 
states [8] that (bosonic) internal symmetries and geometric symmetries
decouple, 
in obvious conflict with (a). However, as there is no proper scattering 
matrix for the charged states, cf.\ (c), the Coleman--Mandula Theorem does 
not apply here. Hence there is room for Lorentz transformations to 
act non--trivially on internal symmetries. 

Secondly, as the charged states cannot be localized adequately, cf.\ (b), 
a proper definition of statistics of the corresponding sectors seems 
impossible. The loopholes are not so obvious here. The essential point
is that, because of (a), any charged state singles out a 
Lorentz frame. Hence one must expect observations in that frame to play 
a special role. Phrased differently, 
some distinguished set of observables should allow one to
identify that particular frame. Obviously, this set of
observables cannot be stable under Lorentz transformations.

The following hypothesis seems attractive: a charged state singles out 
the observables on which it is well localized. These distinguished 
observables include (tensorial) charge densities in some specific 
Lorentz frame. Indeed, in the Coulomb gauge of quantum electrodynamics 
the charged Fermi field commutes at equal times and 
finite distances with the charge density and the magnetic field, but not
with the spatial components of the current 
and the electric field [9]. So there is some evidence that the electric 
charge is, in this restricted sense, localizable; further
support for this idea will be provided by our model. The effects of 
exchanging charges localized in disjoint regions can still be analyzed 
with this restricted notion, thereby providing a basis
for discussing statistics.

The idea that electric charges can be localized in the
restricted sense indicated above was first put forward by Fr\"ohlich
[10], who based a general analysis of superselection sectors
in quantum electrodynamics on this assumption. Yet the problem of
statistics was not discussed in that article. Moreover, some of the
technical assumptions made in the analysis seem physically
unreasonable, cf.\ the remarks below. To ensure a
consistent setting, the present article focuses on a
concrete model illustrating the expected subtle features of the
theories of interest here.

For the convenience of the reader unfamiliar with the
algebraic setting of quantum field theory [11] and the more
recent developments in the theory of superselection sectors
[7] we use the remainder of this introduction to describe
our model in basic field--theoretic terms commenting on the
significance of various steps in our analysis. The expert reader might
wish to proceed directly to the subsequent section.

For our model we draw on the theory of a free, massless, scalar field
$\phi$ in $s+1$ spacetime dimensions in the vacuum Hilbert (Fock)
space ${\Cal H}_0$.  This requires some comment: we are interested in 
long range effects mediated by low energy excitations but believe that
interactions can be neglected without distorting the general qualitative
picture since, according to well--known low energy 
theorems, low energy excitations do not interact with each other
[12].               
However, we use a scalar
field $\phi$ merely as a matter of notational convenience.  Similar models
can be based on the free electromagnetic field. 

The free, massless, scalar field $\phi$ is known to have many
superselection sectors with different infrared properties. We are
interested here in sectors distinguished by a tensorial
charge, cf.\ (a), where Gauss' law holds. The simplest
tensor field constructed from $\phi$ and yielding such a charge 
is 
$$ \Fmnl = ( \dmu g_{ \nu \lambda } - \dnu g_
{ \mu \lambda } ) \Phix,\eqno(1.1)$$ 
where $\dmu$ are
the spacetime derivatives and $g_{ \nu \lambda }$ is the metric. Since
$ \square \phi = 0$, the corresponding identically conserved current is
given by 
$$ j_{ \mu \, ( \lambda ) } (x) = \dno \Fmnl =
\dmu \dla {\Phix}. \eqno(1.2)$$ 
Because of Gauss' law
and locality of the field $\phi$, the resulting charge
operators $Q_{( \lambda )}$ are clearly 0 on the vacuum Hilbert space 
${\Cal H}_0$.

To describe charged states, we have to change the
representation of the algebra generated by the field $\phi$. This can
most easily be done with the help of automorphisms $\gamma$ acting on
polynomials in the field $\phi$. We put 
$$ \gamma (\phi(x) ) := \phi(x) + \varphi(x) \, 1, \eqno(1.3) $$ 
where $\varphi$ is any real distributional solution of the wave
equation. The action of $\gamma$ on arbitrary polynomials in $\phi$ is
obtained from (1.3) by linearity and multiplicativity. In
particular, $\gamma$ acts as follows on charge densities:
$$ \gamma ( \jolx ) = \jolx + \dz \dla \phix \,1. \eqno(1.4) $$ 
Let us now pursue the idea that the
charge density of the states described by $\gamma$ is well
localized. This leads us to take the Cauchy data of $\varphi$ 
to satisfy the following condition: 
$$ \triangle \varphi(0,\vec x) = \rho ( \vec x), \quad
(\dz \varphi ) ( 0, \vec x ) = \sigma ( \vec x), \eqno(1.5) $$
where $\rho, \sigma$ have compact support. 
Then, as $\varphi$ propagates causally, $\gamma$ will act like the
identity on
$\jolx$ for $x$ in the causal complement of a bounded spacetime
region. To simplify the subsequent discussion, we assume that
$\rho, \sigma$ are smooth and compute 
$$ \int d^{ s} \! \vec x \, \dz \dla \phix =\cases  
\int d^{ s} \! \vec x\, \rho ( \vec x ) & \lambda = 0\\ \quad 0 &
\lambda\neq 0.\endcases 
\eqno(1.6) $$ 
It would be premature to infer 
that $\gamma$ describes charged states if $\int d^{ s} \! 
\vec x \, \rho ( \vec x ) \neq 0$, since the operations of integrating
over an 
infinite volume and acting with $\gamma$ cannot simply be interchanged. 
Indeed, a more careful analysis shows
that $\gamma$ does not lead to a new superselection sector if
$s > 3$, i.e., 
the symmetry associated with the current is spontaneously broken in these
cases. Whilst there is an associated Gauss law, the spontaneous breakdown 
is not accompanied by a mass gap in the energy--momentum spectrum, despite 
some of the folklore on the Higgs mechanism. Yet if $s \leq 3$, $\gamma$ 
describes charged states of the charges $Q_{( \lambda )}$ given in (1.6).

If $s \geq 2$, the charged sectors turn out to be invariant
under spacetime translations but of course not under Lorentz
transformations. If $s \geq 3$, translations satisfy the relativistic
spectrum condition. Hence if $s=3$, the model exhibits all
desirable features and we restrict our attention to this case
in the following.

To discuss the statistics of charged sectors,
we have to consider states whose charge densities are localized
in different regions. They are obtained by translating the
automorphisms $\gamma$, setting 
$$ \ga ( \Phix ) := \Phix
+ \varphi ( x \! - \! a ) \, 1 \eqno(1.7) $$
for arbitrary spacetime translations $a$. For the moment let us assume 
for pedagogic reasons that there are unitary
``charged field operators'' $\Wa$, acting on a suitable extension of
${\Cal H}_0$ to a Hilbert space of charged states and implementing the
automorphisms $\ga$, 
$$ \ga ( \Phix ) = \Wai \Phix \Wa. \eqno(1.8) $$ 
(It is not difficult to see that such
operators exist in the present model in certain 
non--regular representations of some Weyl--algebra, cf.\
[13]. We will elaborate on this in the main text.) The
statistics of the charged sectors can then be read off from the
behaviour of their commutators, or better from
their group--theoretic commutator $\eab = \Wai \Wbi \Wa \Wb$, in the
limit of large spacelike separation of $a$ and $b$.

In fact, the statistics operator $\eab$ can be computed without first
having
to solve the more difficult problem of constructing charged
field operators and this is important for a general structural analysis, 
cf.\ the case of strictly localized charges
[1]. To indicate how this can be done and the
type of problems that arise, let us rewrite $\eab$ in
the form 
$$ \eab = ( \Wai \Wb ) \, \Wbi ( \Wbi \Wa )\Wb. \eqno(1.9) $$ 
The operator $\Uab = \Wai \Wb$ appearing in this expression intertwines
the
automorphisms $\gb$ and $\ga$, i.e., 
$$ \Uab\, \gb ( \Phix ) \, \Uabi = \ga ( \Phix ), \eqno(1.10) $$
and can obviously be interpreted as transporting charge. 
In the present model these
charge transporters are defined on the vacuum Hilbert space ${\Cal
H}_0$, but, in contrast to the case of strictly localizable charges, 
they cannot be approximated by local operators in the norm
topology. Such an assumption was made in [10], but does not seem
consistent with gauge theories.

Returning to (1.9), we see that the expression following $\Uab$ 
involves the adjoint action of $\Wbi$, an action, coinciding with that 
of the automorphism $\gb$ on local 
operators on ${\Cal H}_0$, according to (1.8). Yet here
it acts on $U_{b a} = \Uabi$, so the question arises of whether
the charge--carrying automorphisms can be extended to the charge
transporters. If so, one could represent $\eab$ in the form $\eab =
\Uab \, \gb ( \Uabi )$ and the charged fields would have been completely
replaced by quantities intrinsically
defined on the vacuum sector ${\Cal H}_0$.

The physical idea that $\gb$ arises
by shifting a charge from infinity (where it has no effect) to $b$ 
suggests extending the automorphism $\gb$ by setting:
$$ \gb ( U ) := \lim_c \, U_{b c}^{ } U U_{b c}^{-1},
\eqno(1.11) $$  
as $c$ tends to spacelike infinity. So
the question arises of whether this limit exists (in norm) for
all charge transporters $U$, independent of the choice of the
sequence $c$. As the charge transporters factorize:  $U_{b c}
\, U_{c d} = U_{b d}$, the answer to this question is encoded in their
asymptotic commutation properties.
In the model at hand, the relevant conditions are satisfied, enabling 
one to define $\eab$, the
first important step in discussing statistics.

The next problem is whether $\eab$ converges to a limit independent of $a$ 
as $b$ tends spacelike to infinity, as expected if there are charged
fields 
with definite asymptotic commutation relations. What matters here is how 
the charged automorphisms act on intertwiners transporting charges between
distant regions. If this action becomes trivial,
$$ \lim_{c,d} \, \big( \gb ( \Ucd ) - \Ucd \big) = 0 \eqno(1.12) $$
as $c,d$ tend  spacelike to infinity, then the above
limit of $\eab$ exists and has the desired properties.
This condition turns out to be satisfied in the present model. 

To determine the nature of the statistics of the charged sectors, one 
further step is needed. Instead of sending $b$ in $\eab$  spacelike to 
infinity, keeping $a$ fixed, one can interchange the role of 
$a$ and $b$. If the limit is the same, the 
charged sectors have permutation group statistics (and not just braid 
group statistics). With simple sectors, generated by 
automorphisms, as here, the only remaining possibilities are Bose and
Fermi 
statistics. An explicit computation in our model shows that 
$\lim_a \, \eab = \lim_b \, \eab = 1$, implying Bose statistics.\smallskip  

This outline of our results makes it clear that the relevant objects when 
discussing statistics are the charged automorphisms and their intertwiners 
(charge transporters), just as in the case of localized charges. The 
crucial additional information is the 
asymptotic commutation properties of the intertwiners, cf.\ relations 
(1.11) and (1.12). As a matter of fact these properties even suffice 
to go further and to establish systematically that 
charged field operators and a global gauge group exist
in our model. The natural 
mathematical setting for discussing these issues is the theory of 
tensor categories, developed in [14]. We will therefore use 
that setting in the main text. 

\heading
2 The Model
\endheading

   The aim of this section is to give a precise definition of our model. 
It will then be analyzed in subsequent sections and finally
we will comment on those aspects which transcend the particular features 
of our model.\smallskip

   Our model is defined in terms of Weyl operators and the sectors 
will be constructed using automorphisms generated by Weyl operators. 
\smallskip   

   We use standard notation for the Weyl commutation relations: 
$$W(f)W(f')=e^{{i\over 2}\sigma(f,f')}W(f+f'),\eqno(2.1)$$ 
where $\sigma$ is a symplectic form. Our model is just the free massless 
scalar field in $s$ space dimensions, $s\geq 2$.  But we will now confine
our 
attention to the case $s=3$ as we have  already commented 
in the introduction on how the relevant features of the model depend on
$s$. 
Expressing the field in terms of the Cauchy data at time $t=0$, we may
take 
the underlying space to be 
$$\Cal L:= {\omega}^{-{1\over 2}}\Cal D(\Bbb R^s)+i\omega^{1\over 2}\Cal
D(\Bbb R^s).\eqno(2.2)$$ 
Here $\Cal D(\Bbb R^s)$ denotes the space of smooth, real--valued 
functions with compact support and $\omega$ the energy operator. The
symplectic form is given by 
$$\sigma(f,f')=-\text{Im}\langle f,f'\rangle,\eqno(2.3)$$ 
where $\langle\cdot,\cdot\rangle$ is the scalar product on $\Cal L$ 
determining the usual vacuum state, i.e. the one with vanishing one--point 
functional: 
$$\langle f,f'\rangle :=\int d^s\vec x\,\overline{f(\vec x)}f'(\vec
x).\eqno(2.4)$$
The resulting net of von Neumann algebras in the vacuum representation 
will be denoted by $\frak A$.\smallskip

   Our charges will turn out to be localized on the subnet $\frak A_0$ 
which we specify by giving the appropriate subspace $\Cal L_0$ of $\Cal
L$: 
$$\Cal L_0:=\omega^{3\over 2}\Cal D(\Bbb R^s)+i\omega^{1\over 2}\Cal
D(\Bbb R^s).\eqno(2.5)$$ 
Notice that these functions are less singular at the origin in momentum
space. 
By contrast, to define the automorphisms giving rise to the sectors in the 
model, we choose a space $\Cal L_\Gamma$ of functions more singular at the 
origin:
$$\Cal L_\Gamma:=\omega^{-{1\over 2}}\Cal D(\Bbb R^s)+i\omega^{-{3\over
2}}\Cal D(\Bbb R^s).\eqno(2.6)$$ 

   Since $\Cal L$ is invariant under spacetime translations and $\Cal L_0=
i\omega\Cal L$, $\Cal L_0$ is also invariant under spacetime translations. 
It is not, however, invariant under Lorentz boosts. In fact, 
multiplying by $i\omega$ is an isomorphism from $\Cal L$ to $\Cal L_0$ 
considered in the obvious way as nets at time $t=0$. This 
isomorphism induces an isomorphism of nets of von Neumann algebras.
Similarly, since 
$\Cal L_\Gamma=i\omega^{-1}\Cal L$, $\Cal L_\Gamma$ is invariant under 
spacetime translations.\smallskip 

Of course, $\Cal L_\Gamma$ only specifies automorphisms of $\frak A$ if
our symplectic form has been extended in the first variable to $\Cal
L_\Gamma$. 
In fact, our symplectic form $\sigma$ extends in a natural way to 
$\Cal L_\Gamma$. In terms of the smooth functions $g,\,h$ 
parameterizing $\gamma$, 
$$\gamma=i\omega^{-{3\over2}}g+\omega^{-{1\over 2}}h,\quad g,h\in\Cal
D(\Bbb R^s),\eqno(2.7)$$ 
we have 
$$\sigma(\gamma,\gamma')=\int d^s\vec p\,\omega^{-2}(\tilde g(-\vec
p)\tilde h'(\vec p)
-\tilde g'(-\vec p)\tilde h(\vec p)).\eqno(2.8)$$ 
Seen from the point of view of a Weyl algebra, extending to $\Cal
L_\Gamma$ 
in this way would give a Weyl algebra with a non--regular vacuum state,
cf. 
[13]. 
It proves convenient to use the symbol $\gamma$ to denote the automorphism 
generated by $\gamma\in\Cal L_\Gamma$ so that 
$$\gamma(W(f))=e^{-i\text{Im}\langle\omega^{1\over
2}\gamma,\omega^{-{1\over 2}}
f\rangle}\,W(f).\eqno(2.9)$$ 
$\Gamma$ will denote the corresponding group of automorphisms. 
Notice that if $\gamma=\omega^{-{1\over 2}}h$, $h\in\Cal D(\Bbb R^s)$ then 
$\gamma\in L^2(\Bbb R^s)$ so that such $\gamma$ will not 
lead to a new sector. The automorphisms 
in question extend to the local von Neumann algebras. As $s$ is odd, this 
follows from Huygens principle, for 
$\sigma(e^{it\omega}i\omega\gamma,f)=0$ for $t>t_0$, where $t_0$ depends 
only on supp$f$. Thus we have 
$$\sigma(\gamma,f)=-\int_0^{t_0}dt\,\sigma(e^{it\omega}i\omega\gamma,f)\eqno(2.10)$$ 
and $\int_0^{t_0}dt\,e^{it\omega}i\omega\gamma\in L^2(\Bbb
R^s)$.\smallskip

   We regard the automorphisms in $\Gamma$ as defining representations of 
the observable net $\frak A$. These representations are covariant and 
satisfy the spectrum condition. Their unitary equivalence classes will 
be our sectors. For this reason, we define two 
automorphisms $\gamma,\,\gamma'\in\Gamma$ to be equivalent if $\gamma'
\gamma^{-1}$ is induced by a unitary on the vacuum Hilbert space. This is 
equivalent to saying that $\gamma'-\gamma\in\Cal L_\Gamma$ is in the 
$1$--particle space $L^2(\Bbb R^s)$. In terms of the 
smooth functions $g,h$ and $g',h'$ parametrizing $\gamma$ and $\gamma'$ 
we have equivalence if and only if $g-g'$ has a zero at zero in 
momentum space, i.e. if and only if 
$$\int g\,d^s\vec x=\int g'\,d^s\vec x.\eqno(2.11)$$ 

   For us the important property of these automorphisms is that they are 
localized on $\frak A_0$ considered as a net over double cones with axis 
in the time--direction. It suffices to verify that they are localized in 
double cones at time $t=0$, and, because of the causal propagation
properties, 
it suffices to show that given $\gamma\in\Gamma$, there is a ball $B$ at 
time $t=0$ such that
$$\gamma(W(f))=W(f),\eqno(2.12)$$ 
whenever the smooth functions determining $f\in\Cal L_0$ have support in 
the complement of $B$. 
But this follows from 
$$\sigma(\gamma,f)=\sigma(i\omega\gamma,i\omega^{-1}f).\eqno(2.13)$$
\smallskip

   Another property of interest is {\it translatability}. An automorphism 
$\gamma$ of $\frak A$ is said to be translatable if it is equivalent to 
its translates, i.e. if $\gamma_a:=\alpha_a\gamma\alpha_{-a}$ and $\gamma$ 
are equivalent for each spacetime 
translation $a$. Translating $\gamma$ corresponds to translating $g$ and
$h$ 
so translatability follows from the criterion for equivalence derived
above.
As we can dilate the Cauchy data without changing the equivalence class, 
our charged automorphisms are
{\it transportable} in the sense that there are equivalent automorphisms
localized over $\frak A_0$ in any double cone with axis in the
time--direction. In fact, as $s=3$, this is even achieved by the canonical 
automorphic action of the dilation group on $\frak A$ since the induced  
action on $\Gamma$ is given by:
$$g_\lambda(\vec x):=\lambda^{-{s+3\over 2}}\,g(\lambda^{-1}\vec x),
\quad h_\lambda(\vec x):=
\lambda^{-{s+1\over 2}} h(\lambda^{-1}\vec x).\eqno(2.14)$$
Our sectors are thus dilation and translation invariant.

Finally, by virtue of Gauss' law applied to the current (1.2), we
expect to be able to determine the charge carried by the automorphisms
$\gamma$ by testing them with suitable observables, eventually localized
in
the spacelike complement of any given double cone in Minkowski space
${\Bbb R}^{s+1}$. In fact, pick some function
$k\in {\Cal D} ({\Bbb R}^s)$ equal to
$\kappa$ on the unit ball centred at the origin and put
$f_\lambda :=\omega^{3/2} \, k_\lambda, \, \lambda >0$, where
$k_\lambda ( \vec x ):= k ( \lambda^{-1} \vec x )$. $\omega^2\,k_\lambda$ 
has support in the complement of a ball of radius $\lambda$ centred 
at the origin so the Weyl operator $W(f_\lambda)$, which results from 
$W(f_1)$ by the action of dilations, is localized in the spacelike
complement 
of that ball. Hence $W( f_\lambda )$ is a ``central sequence'' commuting
with 
all operators in $\frak A$ in the limit
$\lambda \to \infty$ and the (non--vanishing) vacuum expectation values
$\omega_0 ( W( f_\lambda ))$ do not depend on $\lambda$. Consequently
$$W(f_\lambda) \to \omega_0 (W(f_1)) \, 1 \eqno(2.15)$$
in the weak operator topology in this limit. On the other hand, taking say
$\gamma= i \omega^{-{3\over 2}} g$, with $g\in{\Cal D} ({\Bbb R}^s)$,
we obtain
$$\sigma(\gamma,f_\lambda)=\int\,d^s\vec x \, g(\vec x) k(\lambda^{-1}\vec
x) \eqno(2.16)$$
implying
$$\gamma(W(f_\lambda))=e^{i\sigma (\gamma,f_\lambda)} \, W(f_\lambda)\to
e^{i\kappa q} \, \omega_0(W(f_1)) \, 1\eqno(2.17)$$
in the weak operator topology, where, in virtue of (1.2), 
$q:=\int d^s\vec x \, g$ may be identified as the charge
carried by $\gamma$. Thus inequivalent automorphisms remain inequivalent
on restriction to the observables localized in the spacelike
complement of any double cone. So, as expected, our sectors do
not satisfy the selection criterion of [1].\smallskip

We sum up the results of this section as follows.\smallskip

\noindent
\proclaim{Theorem 1} {\sl The automorphisms in $\Gamma$ are localized
over $\frak A_0$ and transportable forming a group stable under
translations and dilations. The corresponding sectors are invariant
under translations and dilations and do not satisfy the selection
criterion of} [1].
\endproclaim
 
In the next section we consider intertwiners 
between these automorphisms.\smallskip 

\heading 
3 Intertwiners 
\endheading

   Given $\gamma,\,\delta\in\Gamma$, we write $T\in(\gamma,\delta)$ to
denote 
a bounded linear operator on the vacuum Hilbert space such that 
$$T\gamma(A)=\delta(A)T,\quad A\in\frak A.\eqno(3.1)$$ 
$T$ is said to intertwine $\gamma$ and $\delta$. As we are dealing with 
automorphisms and $\frak A$ is irreducible, $(\gamma,\delta)$ is at most 
one dimensional and, if non--zero, consists of multiples of a unitary
operator. 
In this case, $\delta\gamma^{-1}$ is unitarily implementable in the 
vacuum Hilbert space and is hence implemented by a unique Weyl operator 
(see e.g. Lemma A.1 of [15]), namely $W(\delta-\gamma)$.\smallskip

   We regard the set of intertwiners as the arrows in a category, the 
objects then being the automorphisms. Composing arrows means multiplying
the underlying operators and we use the symbol $\circ$ to denote this 
composition whenever emphasis is necessary. 

   We are interested in the asymptotic commutation properties of
intertwiners 
and note that if $U\in(\gamma,\delta)$ and $U'\in(\gamma',\delta')$ are 
non-zero intertwiners then their group--theoretical commutator 
$U{U'}U^{-1}{U'}^{-1}$ is a phase, $\eta(\gamma,\delta;\gamma',\delta')$, 
independent of the choice of $U$ and $U'$. Taking $U$ and $U'$ 
to be Weyl operators, we see that 
$$\eta(\gamma,\delta;\gamma',\delta')=e^{i\sigma(\delta-\gamma,\delta'-\gamma')}.\eqno(3.2)$$
As our symplectic form $\sigma$ has been extended to
$\Cal L_\Gamma$, our phase $\eta$ has a simple combinatorial structure: 
$$\eta(\gamma,\delta;\gamma',\delta')=e^{i\sigma(\delta,\delta')}e^{-i\sigma(\delta,\gamma')}
e^{-i\sigma(\gamma,\delta')}e^{i\sigma(\gamma,\gamma')}.\eqno(3.3)$$\smallskip 

   In the case of localized charges, the intertwiners lie in the
observable net 
as a consequence of duality and the category of intertwiners and
automorphisms 
then acquires the structure of a tensor category. The "tensor product" of 
automorphisms is just composition of automorphisms and needs no special
symbol. 
Given intertwiners $V\in(\gamma,\delta)$ and $V'\in(\gamma',\delta')$
their 
"tensor product" is defined by  
$$V\times V':=V\gamma(V')\in(\gamma\gamma',\delta\delta').\eqno(3.4)$$ 
This is a tensor category since if $U\in(\beta,\gamma)$, and
$U'\in(\beta',\gamma')$ 
then 
$$U\times U'\circ V\times V'=(U\circ V)\times(U'\circ V').\eqno(3.5)$$
Statistics manifests itself as a permutation symmetry $\varepsilon$ for
this 
category. Thus for each pair $\gamma,\gamma'$ of objects we have an 
$\varepsilon(\gamma,\gamma')\in(\gamma\gamma',\gamma'\gamma)$ such that 
$$\varepsilon(\delta,\delta')\circ V\times V'=V'\times
V\circ\varepsilon(\gamma,\gamma'),\quad 
V\in(\gamma,\delta), V'\in(\gamma',\delta'),\eqno(3.6)$$ 
$$\varepsilon(\gamma',\gamma)\circ\varepsilon(\gamma,\gamma')=1_{\gamma\gamma'},\eqno(3.7)$$ 
$$\varepsilon(\gamma\delta,\gamma')=\varepsilon(\gamma,\gamma')\times
1_\delta
\circ1_\gamma\times\varepsilon(\delta,\gamma').\eqno(3.8)$$ 
Sectors described by automorphisms are referred to as {\it simple} sectors 
because the composition law of sectors takes on a simple form making 
the set of sectors into a discrete group whose dual is then a compact
Abelian group 
which is the global gauge group when all sectors are simple sectors.
In general, in superselection theory we have to deal not just with
automorphisms 
but with endomorphisms and their intertwiners and the tensor category has 
a more complicated structure not relevant to our present
discussion.\smallskip

   Returning to our model, although the free massless field 
satisfies duality we cannot conclude that the 
intertwiners lie in the observable net $\frak A$ because our automorphisms 
are localized only relative to $\frak A_0$. Nonetheless, after extending
our 
symplectic form to $\Cal L_\Gamma$, there is an obvious tensor 
category of automorphisms and intertwiners in our case, too. Given 
$U\in (\gamma,\delta)$ and $U'\in (\gamma',\delta')$, we set
$$U\times
U':=U\gamma(U'):=e^{i\sigma(\gamma,\delta'-\gamma')}UU'.\eqno(3.9)$$ 
Together with this tensor product structure there is another phase 
relevant to a discussion of statistics, namely 
$$U\times U'\circ {U'}^{-1}\times U^{-1}=e^{i\sigma(\delta,\delta')}
e^{-i\sigma(\gamma,\gamma')}.\eqno(3.10)$$ 
The form of this equation shows that our tensor category has a permutation 
symmetry $\varepsilon$ given by
$$\varepsilon(\gamma,\gamma')=e^{-i\sigma(\gamma,\gamma')},\eqno(3.11)$$ 
where the phase is understood as a self--intertwiner of $\gamma\gamma'=
\gamma'\gamma$.\smallskip 

   Summing our results up, we conclude:\smallskip 

\noindent 
\proclaim{Theorem 2} {\sl The model has a symmetric tensor category
describing 
a group of simple sectors isomorphic to the discrete group $\Bbb R$. The
objects 
are the elements of $\Gamma$, the arrows are the intertwiners between the 
associated representations of $\frak A$. The tensor structure and the
permutation 
symmetry $\varepsilon$ are determined by the symplectic form
$\sigma$.}\endproclaim
 
If we interpret the permutation symmetry as statistics, 
this means Bose statistics since the phase is one when $\gamma=\gamma'$. 
To understand why this should indeed be regarded as statistics, we must
examine 
the asymptotic behaviour of the symplectic form.\smallskip 

\heading 
4 Asymptotics
\endheading

   We recall that, in the case of strictly localized charges and $s>1$,
intertwiners 
$U\in (\gamma,\delta)$ and $U'\in(\gamma',\delta')$ commute if $\gamma$ 
and $\delta$ are localized in one double cone and $\gamma'$ and $\delta'$
in 
a spacelike separated double cone. We have $U\times U'=U'\times U$ even
under 
the weaker condition that $\gamma$ and $\gamma'$ are spacelike separated 
and $\delta$ and $\delta'$ are spacelike separated. This simple situation 
does not prevail in our model so we investigate the spacelike asymptotic 
behaviour of these commutation properties and, as we have seen, it
suffices to 
consider the behaviour of the symplectic form $\sigma$. Thus we consider
the 
asymptotic dependence of $\sigma(\gamma_a,\gamma'_b)$ on the translations 
$a$ and $b$. Of course, this expression depends only on $a-b$, but we
prefer to 
use the more symmetric form. Writing  
$$\gamma=i\omega^{-{3\over2}}g+\omega^{-{1\over 2}}h,\quad g,h\in\Cal
D(\Bbb R^s),\eqno(4.1)$$ 
we get a sum of four terms: 
$$\sigma(\gamma_a,\gamma'_b)=\sigma(i\omega^{-{3\over2}}g_a,i\omega^{-{3\over2}}g'_b)
+\sigma(i\omega^{-{3\over2}}g_a,\omega^{-{1\over 2}}h'_b)+$$
$$+\sigma(\omega^{-{1\over2}}h_a,i\omega^{-{3\over2}}g'_b)+
\sigma(\omega^{-{1\over 2}}h_a,\omega^{-{1\over 2}}h'_b)\eqno(4.2)$$ 
to be treated separately. The variables involving 
$h$ and $h'$ are in $\Cal L$ so that the final term vanishes exactly
whenever 
the variables have spacelike separated supports. The second term,  
$$\sigma(i\omega^{-{3\over2}}g_a,\omega^{-{1\over 2}}h'_b)=-\int\,d^s\vec
p\,
\omega^{-2}\tilde g(-\vec p)\tilde h'(\vec
p)\,\text{cos}\omega(a^0-b^0)e^{i\vec p
\cdot(\vec a-\vec b)},\eqno(4.3)$$ 
decays like $|\vec a-\vec b|^{-1}$. Set 
$$Z(x):=\int d^s\vec p\,\omega^{-2}\tilde g(-\vec p)\tilde h'(\vec
p)\text{cos}
\omega x^0e^{i\vec p\cdot\vec x},\eqno(4.4)$$ 
then
$$Z(x)=Z(0,\vec x)+\int^{x^0}_0 dt\,\dot Z(t,\vec x).\eqno(4.5)$$ 
Since $\dot Z$ is a solution of the wave equation with compact Cauchy
data, 
the integral does not contribute when $|\vec x|>|x^0|+$ const.\ and the
first 
term can be written:
$$Z(0,\vec x)=\text{const}\,\int d^s\vec y\, (g*h')(\vec y)\,|\vec x-\vec
y|^{2-s}.\eqno(4.6)$$ 
Hence $Z(x_0,\vec x)=\text{O}(|\vec x|^{2-s})$ for $|\vec x|>|x^0|+$
const. 
The term involving $g'$ and $h$ 
is of exactly the same type and thus decays in the same way.\smallskip 

   The remaining term is 
$$\sigma(i\omega^{-{3\over2}}g_a,i\omega^{-{3\over2}}g'_b)=\int\,d^s\vec
p\,
\omega^{-3}\tilde g(-\vec p)\tilde g'(\vec
p)\,\text{sin}\omega(a^0-b^0)e^{-i\vec p
\cdot(\vec a-\vec b)}.\eqno(4.7)$$ 
Setting 
$$Y(x):=\int\,d^s\vec p\,\omega^{-3}\tilde g(-\vec p)\tilde g'(\vec p)
\,\text{sin}\omega x^0 e^{-i\vec p\cdot\vec x},\eqno(4.8)$$ 
and writing 
$$Y(x)=Y(0,\vec x)+x^0\dot Y(0,\vec x)+\int_0^{x^0}dt\,\int_0^tds\,\ddot
Y(s,\vec x),\eqno(4.9)$$ 
we see that the first summand vanishes.  $\ddot Y$ is a solution of the
wave equation 
with compact Cauchy data and hence the integral vanishes for sufficiently
spacelike 
$x$. 
Finally,
$$x^0\,\dot Y(0,\vec x)=c\tilde g(0)\tilde g'(0)\,{{x^0}\over{|\vec
x|^{s-2}}}
+\text O( {{x^0}\over{|\vec x|^{s-1}}}),\eqno(4.10)$$ 
where $c$ is a non-zero constant. 
Summing up what we have learned, we have\smallskip 

\noindent 
\proclaim{Theorem 3} The asymptotic behaviour of the symplectic form for
$s=3$ 
as $a-b$ tends spacelike to infinity is 
$$\sigma(\gamma_a,\gamma'_b)={1\over 4\pi}\,qq'{{|a^0-b^0|}\over{|\vec
a-\vec b|}} +\text o(1).\eqno(4.11)$$ 
Here $q=\int d^s\vec x g$ and $q' = \int d^s\vec x g'$ are the charges
carried
by $\gamma$ and $\gamma'$.
\endproclaim
 
In this way we see that intertwiners will commute asymptotically and the 
cross product will commute asymptotically provided we go spacelike to 
infinity in such a way that 
$${{a^0-b^0}\over{|\vec a-\vec b|}}\to 0.\eqno(4.12)$$ 

\heading 
5 Interpretation 
\endheading

   After the computations on the asymptotic behaviour of the 
symplectic form in the last section, we show how to arrive at the 
tensor category reasoning just in terms of the representations defined by 
the elements of $\Gamma$ and the corresponding intertwiners without 
making use of an ad hoc, if natural, extension of the symplectic form 
from $\Cal L$ to $\Cal L_\Gamma$.\smallskip 

   Given $\gamma,\,\delta,\,\delta'\in \Gamma$ and $V\in
(\delta,\delta')$, 
we wish to give a meaning to $\gamma(V)$ as an intertwiner from 
$(\gamma\delta,\gamma\delta')$. In the case of localized charges it
suffices 
to define 
$$\gamma(V):=U^*VU,\eqno(5.1)$$ 
where $U\in (\gamma,\gamma')$ is unitary and $\gamma'$ is localized
spacelike to $V$. 
The physical idea behind this construction is to regard $U$ as an
operation 
of transferring charge. Thus the formula: 
$$\gamma(V):=\lim_aU_a^*VU_a,\quad U_a\in(\gamma,\gamma_a),\eqno(5.2)$$ 
where the limit is taken as $a$ tends spacelike to infinity, would be
valid 
for any intertwiner $V$ and we have created the charge carried by $\gamma$ 
by transferring it from spacelike infinity.\smallskip 

   We will use the same formula to define $\gamma(V)$, but now requiring 
that ${a^0\over{|\vec a|}}\to 0$. In fact, in our model, 
$$U_a^*VU_a=\eta(\delta,\delta';\gamma,\gamma_a)V\eqno(5.3)$$ 
and expressing $\eta$ in terms of the symplectic form $\sigma$ and using 
its asymptotic behaviour, we conclude that 
$$U_a^*VU_a\to e^{i\sigma(\delta-\delta',\gamma)}V.\eqno(5.4)$$ 
But this means that our new definition of $\gamma(V)$ coincides with the 
old definition and hence serves to turn our category of automorphisms and 
intertwiners into a tensor category.\smallskip 

   The same line of reasoning leads us to a definition of statistics.
We know that in the case of strictly localized charges the permutation 
symmetry $\varepsilon$ describing the statistics can be characterized by 
requiring that
$$\varepsilon(\gamma,\delta)=(V\times U)^{-1}\circ (U\times
V),\eqno(5.5)$$ 
when $U\in(\gamma,\gamma')$, $V\in(\delta,\delta')$ are unitaries and 
$\gamma'$ and $\delta'$ are spacelike separated. Hence we can try 
defining the statistics operator by 
$$\varepsilon(\gamma,\delta)=\lim_a\,(V_b\times U_{a})^{-1}\circ 
U_{a}\times V_b,\eqno(5.6)$$     
where $U_a\in(\gamma,\gamma_a)$ and $V_b\in(\delta,\delta_b)$ are unitary. 
If we compute in our model, we find 
$$\varepsilon(\gamma,\delta)=\lim_a e^{i\sigma(\delta,\gamma)}e^{-i\sigma(
\delta_b,\gamma_{a})}.\eqno(5.7)$$ 
Using the asymptotic properties of the symplectic form, we find 
$$\varepsilon(\gamma,\delta)=e^{-i\sigma(\gamma,\delta)}\eqno(5.8)$$ 
Note that we might equally well have kept $a$ fixed and sent $b$ to 
spacelike infinity or have sent $a$ and $b$ to spacelike infinity in 
opposite directions. So we have verified that the symmetric tensor
category  
and hence, in particular, the statistics, may be derived by physically 
motivated conditions of an asymptotic nature.\smallskip 

   The importance of being able to describe the superselection structure
in 
terms of a symmetric tensor category is that it opens the way to
constructing 
a field net with normal commutation relations acted on by a compact gauge 
group whose fixed--point net is the observable net. This construction was 
first carried out for simple sectors in [16] 
and in the case of general localized charges in [7]. We do not need 
to discuss how the scheme should be modified to embrace the setting of our 
model since we get the field net as the net of von Neumann algebras 
in the vacuum sector generated by $\Cal L_\Gamma$ with the extended 
symplectic form.\smallskip

\heading 
6 Outlook
\endheading 

At the first sight, it might appear that our model has many special
features 
rendering the above analysis possible. For example, we were able to 
verify the existence of norm limits by examining the behaviour of certain
phases. 
However, the intertwining spaces will be one--dimensional whenever we look
at 
sectors generated by automorphisms. Although these automorphisms will not 
commute, in general, as automorphisms generated by Weyl operators do, they 
may have sufficient commutativity properties. In the case of localized 
charges, automorphisms localized spacelike to one another commute and
there 
are generalizations to the case of localization relative to a suitably 
large subnet. In these cases, our phases will be defined whenever the 
automorphisms in question have the appropriate spacelike localization 
properties and the asymptotic norm convergence reduces to the convergence 
of these phases, just as in the model. If we do have convergence 
to appropriate limits then we get a symmetric tensor category of 
automorphisms and intertwiners describing the corresponding superselection 
structure. This will be discussed elsewhere.\smallskip 

Thus from a technical point of view we might say that, although our model 
has simplifying features allowing an easy complete analysis of the 
sectors in question, the analysis itself applies to a whole class of
theories.
It is not clear how many physically interesting theories would fall 
within this class. Yet there are many models having charges localized on
some 
large subnet and it would seem that electric and magnetic charges 
should exhibit this type of behaviour. 
\smallskip 
 
\heading
References
\endheading

\item{[1]} S.\ Doplicher, R.\ Haag and J.E.\ Roberts {\it Local 
Observables and Particle Statistics, I\/}, Commun.\ Math.\ Phys.\ {\bf 23} 
(1971), 199--230, and {\it II\/}, Commun.\ Math.\ Phys.\ {\bf 35} (1974), 
49--85 .

\item{[2]} D.\ Buchholz and K.\ Fredenhagen {\it Locality and the 
Structure of Particle States\/}, Commun.\ Math.\ Phys.\ {\bf 84} (1982), 
1--54. 

\item{[3]} J.\ Fr\"ohlich, G.\ Morchio and F.\ Strocchi {\it 
Charged Sectors and Scattering States in Quantum Electrodynamics\/}, 
Ann.\ Phys.\ {\bf 119} (1979), 241-284.

\item{[4]} D.\ Buchholz {\it Gauss' Law and the Infraparticle Problem\/}, 
Phys.\ Lett.\ {\bf B174} (1986), 331-334. 

\item{[5]} F.\ Strocchi {\it Selected Topics on the General Properties 
of Quantum Field Theory}, World Scientific 1993.

\item{[6]} D.\ Buchholz {\it The Physical State Space of Quantum 
Electrodynamics\/}, Commun.\ Math.\ Phys.\ {\bf 85} (1982), 49--71.

\item{[7]} S.\ Doplicher and J.E.\ Roberts {\it Why there is a Field 
Algebra with a Compact Gauge Group Describing the Superselection Structure 
in Particle Physics\/}, Commun.\ Math.\ Phys.\ {\bf 131} (1990), 51--107.

\item{[8]} J.\ Lopuszanski {\it An Introduction to Symmetry and
Supersymmetry
  in Quantum Field Theory\/}  World Scientific: Singapore 1991

\item{[9]} K.\ Symanzik {\it Lectures on Lagrangian Quantum Field
Theory\/},
Desy Report T-71/1.

\item{[10]} J.\ Fr\"ohlich {\it The Charged Sectors of Quantum
Electrodynamics 
in a Framework of Local Observables\/}, Commun. Math. Phys. {\bf 66}
(1979), 223-265.

\item{[11]} R.\ Haag {\it Local Quantum Physics\/}, Springer: Berlin, 
Heidelberg, New York 1992. 

\item{{[12]}} J.\ M.\ Jauch, F.\ Rohrlich {\it The Theory of Photons 
and Electrons\/} Springer: Heidelberg 1980.

\item{[13]} F.\ Acerbi, G.\ Morchio and F.\ Strocchi {\it Theta Vacua, 
Charge Confinement and Charge Sectors from Nonregular Representations of 
CCR Algebras} Lett.\ Math.\ Phys.\ {\bf 27} (1993), 1--11; {\it Infrared 
Singular Fields and Non--regular Representations of CCR Algebras\/}, 
Jour.\ Math.\ Phys.\ {\bf 34} (1993), 899--914.

\item{[14]} S.\ Doplicher and J.E.\ Roberts {\it A New Duality Theory 
for Compact Groups\/}, Inventiones Math.\ {\bf 98} (1989), 157--218.

\item{[15]} D.\ Buchholz, S.\ Doplicher, R.\ Longo, J.E.\ Roberts:
{\it A New Look at Goldstone's Theorem}, Rev.\ Math.\ Phys.\ Special 
Issue (1992), 49-83.

\item{[16]} S.\ Doplicher, R.\ Haag, J.E.\ Roberts: {\it Fields, 
Observables and Gauge Transformations II}, Commun.\ Math. Phys.\ 
{\bf 15} (1969), 173-200.

\enddocument
\bye